\documentclass[a4paper,12pt]{article} 
\usepackage[T1]{fontenc}
\usepackage{amsfonts,amsmath,amssymb,enumerate,epsfig,graphics}

\numberwithin{equation}{section}

\newtheorem{definition}{Definition}[section]
\newtheorem{proposition}[definition]{Proposition}

\newcommand{\prf}{\underline{Proof:}\ }
\newcommand{\finprf}{\null \hfill {\rule{5pt}{5pt}}\\ \null}

\newcommand{\ie}{{\it i.e.}\ }
\newcommand{\be}{\begin{equation}}
\newcommand{\ee}{\end{equation}} 
\newcommand{\bma}{\begin{pmatrix}}
\newcommand{\ema}{\end{pmatrix}}
\newcommand{\bea}{\begin{eqnarray}}
\newcommand{\eea}{\end{eqnarray}}

\newcommand{\RR}{\mbox{${\mathbb R}$}}
\newcommand{\CC}{\mbox{${\mathbb C}$}}

\newcommand{\cB}{\mbox{${\cal B}$}}
\newcommand{\cL}{\mbox{${\cal L}$}}

\makeatletter

\let\captionORI\caption
\def\caption#1{\captionORI{\rm\small #1}}
\makeatother

\begin{document}

\pagestyle{empty}
\setcounter{page}{0}
\null

\begin{center}

{\LARGE \bf Vector Nonlinear Schr\"odinger Equation\\
 on the half-line}

\vspace{1cm}

{\large V. Caudrelier and Q. C. Zhang }\\

\vspace{0.5cm}

{\it Centre for Mathematical Science,
City University London, \\
Northampton Square,London EC1V 0HB, UK.}
\end{center}

\vfill

\begin{abstract}
We investigate the Manakov model or, more generally, the vector nonlinear Schr\"odinger equation on the half-line. 
Using a B\"acklund transformation method, 
two classes of integrable boundary conditions are derived: mixed Neumann/Dirichlet and Robin boundary conditions. 
Integrability is shown by constructing a generating function for the 
conserved quantities. We apply a nonlinear mirror image technique to construct the inverse scattering method 
with these boundary conditions. The important feature in the reconstruction formula for the 
fields is the symmetry property of the scattering data emerging from the presence of the boundary. Particular attention is paid to the discrete spectrum. 
An interesting phenomenon of transmission between the components of a vector soliton interacting with the boundary is demonstrated. This is specific to the 
vector nature of the model and is absent in the scalar case. For one-soliton solutions, we show that the boundary can be used to make
certain components of the incoming soliton vanishingly small. This is reminiscent of the phenomenon of light polarization by reflection.
\end{abstract}
PACS numbers: 02.30.Ik, 02.30.Zz.\\
AMS classification (2010): 35Q55, 37K10, 37K15.

\vfill


\newpage
\pagestyle{plain}

\section{Introduction}\label{sec:1}

In this paper, we consider the vector nonlinear Schr\"odinger equation (VNLS) on the half-line. Historically, the model on the line for a vector field 
with two components was introduced and solved by the inverse scattering method by Manakov \cite{Man}. The VNLS on the half-line can be seen as a generalization 
of the famous (scalar) one-dimensional nonlinear Schr\"odinger equation (NLS)
along two directions: adding internal degrees of freedom and adding a boundary. It is worth mentioning that each direction is not 
just a trivial generalization exercise. Indeed, such considerations have been at the origin of major breakthroughs in the history of integrable systems.
For instance, when the original Lieb-Liniger model \cite{LL} for spinless particles with contact interactions was generalized by Yang \cite{Yang} 
to particles with 
spin, this gave rise to the now famous rational solution of the Yang-Baxter equation. Similarly, when studying integrable systems with boundaries, 
Sklyanin unraveled new algebraic structures controlled by the so-called reflection equation \cite{Skly_boundary}. Concerning the NLS equation, 
these questions have been studied rather well at the quantum level. Vector generalizations on the line \cite{MW,MRSZ} and on the half-line \cite{MRS2}
have been investigated essentially from an algebraic point of view. At the classical level, a very general framework for dealing with 
NLS equations related to symmetric spaces has been presented in \cite{FK}. Although the VNLS equation is only a particular case of this general theory, 
it is still the object of active research: for 
instance, the full understanding of the factorization property of vector soliton collisions has only been achieved relatively recently (see for instance 
\cite{Tsu} and references therein). At the classical level again, the 
general question of generating boundary conditions of arbitrary order compatible with a general multicomponent nonlinear Schr\"odinger equation 
(expressed in terms of 
Jordan triple algebras) has been treated in \cite{HS}. However, to the best of the authors' knowledge (and surprise), 
the problem of formulating the inverse scattering method with integrable boundary conditions to tackle for instance the question of 
interactions of vector solitons with a boundary has not been addressed yet. 
This is in sharp contrast with the scalar case where a rich literature has appeared over the years since the early attempts of adapting 
the inverse scattering 
method to initial-boundary value problems \cite{AS} up to the very general method developed by Fokas (see e.g. the textbook
 \cite{fokasbook}), 
in passing by the B\"acklund transformation method found in \cite{H,BT1,BT2}. Recently, another idea \cite{Fokas2} was revived in \cite{BH} and 
used to investigate soliton interactions in the presence of an integrable boundary by means of a (nonlinear) mirror image technique. As far as 
integrable boundary conditions are concerned, all these techniques are related and it seems that the central object is a special B\"acklund 
matrix that allows for the implementation of the mirror image technique. In the context of Fokas' method, this matrix evaluated at the 
location of the boundary is the matrix involved in the so-called linearizable boundary conditions.

The goal of this paper is to introduce and study the VNLS with an integrable boundary in order to lay the foundations for a deeper 
understanding of integrable vector field interactions with a boundary. Section \ref{sec:2} contains known results for the 
inverse scattering method of the VNLS on the line. This allows us to introduce our notations as well as to collect useful results for the rest 
of the paper. Section \ref{half-line} represents the core of this paper. We first derive two classes of integrable boundary conditions using the B\"acklund matrix 
approach. Integrability is established by constructing an explicit generating function for the conserved quantities for the VNLS on the half-line. 
This is then used to implement the mirror image technique that allows us to use the inverse scattering method results on the line to 
obtain the reconstruction formulae for the fields on the half-line. In section \ref{examples}, we use our general results to investigate 
the behaviour of one soliton bouncing off a boundary. An interesting phenomenon of transmission between different modes of polarization 
is demonstrated. It is due to the interaction between the soliton polarization and what we can call the natural boundary polarization. This will be 
made more precise in the text. Our conclusion are gathered in section \ref{conclusions}.

The vector nonlinear Schr\"odinger equation (VNLS) on the half line is defined as the following initial-boundary value problem for the $n$-component 
vector field $R(x,t)$

\bea \label{equ:schr}
 i \frac{\partial R}{\partial t} + \frac{ \partial^2 R}{\partial x ^2} - 2 \lambda R R^\dagger R = 0\,, \qquad x, t \in [0, \infty )\,,\\
 R(x,0)=R_0(x)~,~~R(0,t)=g_0(t)~,~~R_x(0,t)=g_1(t)\,,
\eea
where  $R^\dagger(x,t)$ is the transpose conjugate of $R(x,t)$ and $ \lambda $ is the (real) coupling constant which can be normalised to 
$\lambda = \pm 1 $. We assume that the functions $R_0$, $g_0$ and $g_1$ live in appropriate functional spaces so as to ensure that the 
calculations in this paper are meaningful (for details in the scalar case see for instance the "rigorous considerations" section in \cite{FI}). In particular, they are decaying functions at infinity.
A powerful method has been developed by Fokas and various co-workers over the last two decades to tackle integrable nonlinear equations 
on the half-line (see e.g. \cite{fokasbook} and references therein).The key ingredient is the simultaneous treatment of the two equations appearing in the Lax pair formulation 
of the nonlinear equation. We follow this principle in this paper.

\section{Vector NLS on the line}\label{sec:2}

In this section, we review the inverse scattering method for VNLS on the line. We follow chapter four of the textbook \cite{APT} and collect the main results and 
notations needed in the rest of the paper. 

\subsection{Lax pair}\label{sect_Lax}

Given $R (x,t)$, define $Q(x,t)$ as the following $(n+1) \times (n+1)$ matrix-valued field
\be \label{equ:field}
Q(x,t) = \bma 0 & R(x,t) \\ \lambda R^{\dagger}(x,t) & 0 \ema\,.
\ee
Equation (\ref{equ:schr}) is the compatibility condition of the following linear problem for the matrix-valued field 
$\Phi(x,t,k)$\footnote{ From now on, we drop the $x,t$ (and $k$) dependence for conciseness unless there is ambiguity.}
\bea
\label{equ:laxp1} 
  &&\Phi_x + ik [ \Sigma _3 , \Phi] =  Q\,\Phi \,,\\
\label{equ:laxp2} 
  &&\Phi_t + 2 i k^2 [\Sigma_3, \Phi]=   G\, \Phi\,,
\eea
where
\be
\Sigma_3 = \bma I_n & 0 \\ 0 & -1  \ema~~\text{and}~~G = 2k Q -  i Q_x \Sigma_3 -  iQ^2 \Sigma_3\,, 
\ee
$I_n$ being the $n\times n$ identity matrix.
Using standard arguments, one can show that $\det \Phi(x,t,k)=f(k)$ for all $x,t\in\RR$, and all values of $k\in\CC$ for which 
$\Phi$ is defined, where $f$ can be determined for instance by fixing the asymptotic behaviour of $\Phi$ as $x\to\pm\infty$.
Also, if $\Phi$ and $\Psi$ are two solutions of (\ref{equ:laxp1},\ref{equ:laxp2}) then 
\bea
\label{2solutions}
\Phi(x,t,k)=\Psi(x,t,k)\,e^{-i\phi(x,t,k)\Sigma_3} \,T(k)\,e^{i\phi(x,t,k)\Sigma_3}\,,
\eea
where $T$ depends on $k$ only and $\phi (x,t,k) = kx + 2k^2 t$. Finally, from 
\be
\label{equ:inv}
W Q W^{-1} = - Q^{\dagger}\,,
\ee
where 
\be
W = \bma -\lambda I_n & 0 \\ 0 & 1 \ema\,,
\ee
one obtains that $W \Phi^{\dagger} (x,t,k^*) W^{-1}$ and $\Phi^{-1} (x,t,k)$ satisfy the same equations and are thus related as in (\ref{2solutions}).

\subsection{Direct scattering problem}

We define two fundamental (or Jost) solutions $M_\pm$ of (\ref{equ:laxp1},\ref{equ:laxp2}) satisfying
\bea
\label{asymptotics}
\lim_{x\to\pm\infty}e^{i\phi(x,t,k)\Sigma_3 }M_\pm(x,t,k)e^{-i\phi(x,t,k)\Sigma_3 }=I_{n+1}~~,~~k\in\RR\,.
\eea
They have the following Volterra integral equation representations
\be
\displaystyle  \label{equ:volte}
M_\pm(x,t,k)=I_{n+1} + \int_{\pm\infty}^{x} e^{ - i k(x-y)\Sigma_3 } Q(y,t)M_\pm(y,t,k)e^{ik(x-y)\Sigma_3 }\,dy\,.
\ee
Their analyticity and boundedness properties then follow as in the scalar case. Splitting the matrices into left and right "column vectors"
\footnote{Here, the left column vector is made of the first $n$ left columns and the right one is made of the remaining column.}, one can 
indicate the domain of the complex $k$ plane where they are bounded and analytic using the following notations
\be
M_+(x,t,k) = (\overline{N}(x,t,k),N(x,t,k))\,,~~M_-(x,t,k)=(M(x,t,k),\overline{M}(x,t,k))\,,
\ee
where $\overline{N}$, $\overline{M}$ (resp. $N$, $M$) are bounded analytic in the lower (resp. upper) half plane as functions of $k$.
From (\ref{asymptotics}), one finds that 
\bea
\det M_\pm(x,t,k)=1\,.
\eea
Having defined the fundamental solutions, one can proceed with the definition of the scattering data. Recalling (\ref{2solutions}), we define
the matrix $S(k)$ for $k\in\RR$ as
\bea
\label{def_S}
e^{i\phi(x,t,k)\Sigma_3} M_-(x,t,k) e^{-i\phi(x,t,k)\Sigma_3}=e^{i\phi(x,t,k)\Sigma_3} M_+(x,t,k) e^{-i\phi(x,t,k)\Sigma_3}\, S(k)\,.
\eea
Several important properties follow from this definition. First, $\det S(k)=1$. Then, $S(k)$ has the following integral representation
\bea
\label{rep_S}
S(k)=I_{n+1} + \int_{-\infty}^{\infty} e^{ i ky\Sigma_3 } Q(y,0)M_-(y,0,k)e^{-iky\Sigma_3 }\,dy\,.
\eea
Splitting $S(k)$ into block matrices of natural sizes\footnote{For instance, $a$ is $n\times n$ and $\bar{a}$ is a scalar.}
\bea
S(k) = \left(\begin{array}{cc}
a(k)&\bar{b}(k)\\
b(k)&\bar{a}(k)
\end{array}\right)\,,
\eea 
one gets that $a(k)$ can be analytically continued to $\text{Im}\,k>0$ while $\bar{a}(k)$ can be analytically continued to $\text{Im}\,k<0$. It is known 
that $\bar{a}(k^*)=(\det a(k))^*$. In the focusing case ($\lambda=-1$), the usual assumption is that $\det a(k)$ has a finite number of simple zeros located in the upper half-plane off 
the real axis. Denote them $k_j$, $\text{Im} \,k_j>0$, $j=1,\dots,J$. Consequently, $\bar{a}(k)$ has the same number $J$ of simple zeros 
$k_j^*$, $j=1,\dots,J$ in the lower half-plane. These zeros play an important role in the inverse scattering problem described below. In particular, 
the residues of $b(k)a^{-1}(k)$ at $k=k_j$, called norming constants, are used to reconstruct the solitonic part of the vector field $R$.

\subsection{Inverse scattering problem}\label{ISP}
Let us rewrite the relations (\ref{def_S}) as
\bea
M(x,t,k)a^{-1}(k)&=&\overline{N}(x,t,k)+e^{2i\phi(x,t,k)}N(x,t,k)\rho(k)\,,\\
\overline{M}(x,t,k)\bar{a}^{-1}(k)&=&e^{-2i\phi(x,t,k)}\overline{N}(x,t,k)\bar{\rho}(k)+N(x,t,k)\,,
\eea
where $\rho(k)=b(k)a^{-1}(k)$ and $\bar{\rho}(k)=\bar{b}(k)\bar{a}^{-1}(k)$. Next, apply the 
Cauchy projector $P^-$ (resp. $P^+$) to the first (resp. second) equality where
\bea
P^\pm(f)(k)=\frac{1}{2i\pi}\int_{-\infty}^\infty\frac{f(\omega)}{\omega-(k\pm i0)}\,d\omega\,.
\eea
Using standard contour integration techniques and the asymptotic behaviour in $k$ that can be obtained using integration by parts on the integral 
representations (\ref{equ:volte}) and (\ref{rep_S}), one obtains 
\bea
\label{coupled_eq1}
\overline{N}(x,t,k)=\bma
I_n\\
0
\ema&+&
\sum_{j=1}^J \frac{e^{2ik_jx+4ik_j^2t}}{k-k_j}N(x,t,k_j)\,C_j\nonumber\\
&+&
\frac{1}{2i\pi}\int_{-\infty}^\infty\frac{e^{2i\omega x+4i\omega^2t}N(x,t,\omega)\rho(\omega)}{\omega-(k- i0)}\,d\omega\,,\\
\label{coupled_eq2}
N(x,t,k)=\bma
0\\
1
\ema&+&
\sum_{j=1}^J \frac{e^{-2ik_j^*x-4ik_j^{*2}t}}{k-k_j^*}\overline{N}(x,t,k_j^*)\,\overline{C}_j\nonumber\\
&-&
\frac{1}{2i\pi}\int_{-\infty}^\infty\frac{e^{-2i\omega x-4i\omega^2t}\overline{N}(x,t,\omega)\bar{\rho}(\omega)}{\omega-(k+ i0)}\,d\omega\,,
\eea
where 
\bea
C_j=Res(b(k)a^{-1}(k),k_j)~~\text{and}~~\overline{C}_j=Res(\bar{b}(k)\bar{a}^{-1}(k),k_j^*)\,.
\eea
Note that (\ref{equ:inv}) implies that $\overline{C}_j=\lambda C_j^\dagger$.
Evaluating the first equation at $k=k_j^*$ and the second at $k=k_j$, one obtains a closed system of linear algebraic integral equations that 
can be solved in principle\footnote{That this is the case is usually established by using the equivalent formulation in terms of a Riemann-Hilbert 
problem.}. Below, we will see that this system of equations allows for closed-form formulas
 in the purely solitonic case \ie when one sets $\rho(k)=0=\bar{\rho}(k)$ for $k\in\RR$. 
 
 In the scalar case, the norming constants $\gamma_\ell$ are defined as the proportionality constants between the two column vector forming a Jost 
 solution when the spectral parameter equals to a zero $k_\ell$
 \bea
 M(x,t,k_\ell)=e^{2i\phi(x,t,k_\ell)}N(x,t,k_\ell)\gamma_\ell\,.
 \eea
 When working in the class of exponentially fast decreasing functions, there is a nice characterization of $\gamma_\ell$ in terms of $b(k)$ which 
 can be analytically continued up to the strip $\{k\in\CC ; 0\le \text{Im} \,k\le K\}$ where $K$ controls the decrease of the functions. 
 With $K\ge \text{max}\{\text{Im}\, k_j;j=1,\dots,J\}$, we then have $\gamma_\ell=b(k_\ell)$.
 In the vector case, the norming constants can be defined as the vectors describing the linear dependence of the column of the Jost solution at the 
 zeros $k_j$
 \bea
  M(x,t,k_j)\Gamma_j=e^{2i\phi(x,t,k_j)}N(x,t,k_j)C_j\,,
 \eea
 where $\Gamma_j$ is an $n\times n$ matrix and $C_j$ a row vector. Again, in the class of exponentially fast decreasing functions one has, without
 loss of generality,
 \bea
 \Gamma_j=A(k_j)^t~~\text{and}~~C_j=\frac{1}{\alpha^{'}(k_j)}b(k_j)A(k_j)^t\,,
 \eea
where $\alpha^{'}(k)=\frac{d\det a(k)}{dk}$ and $A(k)^t$ is the transpose of the cofactor matrix of $a(k)$. Although not the most general, we will work with this 
convenient characterization to derive in particular proposition (\ref{main_result}) below. We stress that the results 
would still be valid in a more general case, but this is enough for our purposes of studying soliton solutions.

 Once $N$ and $\overline{N}$ are known, one obtains $R(x,t)$ by the formula
 \bea
 \label{R_reconstructed}
 R(x,t)=2i\lambda\sum_{j=1}^Je^{-2ik_j^*x-4ik_j^{*2}t}\overline{N}^{(up)}(x,t,k_j^*) C_j^\dagger+
 \frac{1}{\pi}\int_{-\infty}^\infty e^{-2i\omega x-4i\omega^2t}\overline{N}^{(up)}(x,t,\omega)\bar{\rho}(\omega)\,d\omega\,,\nonumber\\
 \eea
 where $\overline{N}^{(up)}$ is the $n\times n$ upper block of $\overline{N}$.
This is obtained by comparing the $O(1/k)$ term in the expansion obtained from (\ref{equ:volte}) by integration by parts 
and the $O(1/k)$ term obtained from (\ref{coupled_eq2}). 

In the pure soliton case, (\ref{coupled_eq1},\ref{coupled_eq2}) yield the following coupled equations
\bea
\label{coupled_eqs1}
\overline{N}(x,t,k_\ell^*)-\lambda\sum_{j,m=1}^J \frac{e^{2i(k_j-k_m^*)x+4i(k_j^2-k_m^{*2})t}}{(k_\ell^*-k_j)(k_j-k_m^*)}N(x,t,k_m^*)C^\dagger_m C_j\,.
=I_n~~,~~\ell=1,\dots,J\,.
\eea
Introducing the matrix $\mu(x,t)$ whose entries are the following $n\times n$ matrices
\bea
\mu_{m\ell}(x,t)=\delta_{m\ell}\,I_n-\lambda\sum_{j=1}^J\frac{e^{2i(k_j-k_m^*)x+4i(k_j^2-k_m^{*2})t}}{(k_\ell^*-k_j)(k_j-k_m^*)}C^\dagger_m C_j\,,
\eea
and the notation $\overline{N}(x,t,k_\ell^*)=\overline{N}_\ell$, equations \eqref{coupled_eqs1} can be compactly rewritten as
\bea
\bma
\overline{N}_1\,\overline{N}_2\,\dots\,\overline{N}_J
\ema\,\mu(x,t)=\bma
I_n\,I_n\dots I_n
\ema\,.
\eea
One then obtains 
\bea
R(x,t)=2i\lambda \bma
I_n\,I_n\dots I_n
\ema\mu^{-1}(x,t)
\bma
C_1^\dagger e^{-2ik_1^*x-4ik_1^{*2}t}\\
\vdots\\
C_J^\dagger e^{-2ik_J^*x-4ik_J^{*2}t}
\ema\,.
\eea

This is a central formula to understand the behaviour of solitons. In the next section, we show 
how to modify this formula to take into account the presence of the boundary. This is then used in Section \ref{examples} for numerical simulations.

\section{Vector NLS on the half-line}\label{half-line}

Although the most general method for tackling initial boundary value problem exactly is the one developed by Fokas and co-workers, 
since we restrict our attention to a family of integrable boundary conditions (the vector generalisation of the Robin boundary conditions), we 
employ the mirror image technique as used in \cite{BH}. These two approaches give equivalent results for this class of boundary 
conditions which would be called linearizable in the terminology used by Fokas. The advantage of the mirror image technique is that it allows one 
to use the results on the line with minor (but important) modifications to get explicit solutions. One inconvenience though is that one has to prove that the boundary conditions are indeed satisfied 
by the field. This is because one uses the framework presented in the previous section for a modified field $P(x,t,k)$ instead 
of $Q(x,t)$ whose structure is chosen a priori. We note that, to the authors' knowledge, such a check has only been performed explicitly in \cite{Fokas2} 
and only in the case of Dirichlet boundary conditions. Below, we present a similar proof for the mixed Neumann/Dirichlet boundary conditions in the 
vector case. We also show that the class of boundary conditions we consider is integrable in the sense that it allows for an infinite number of conserved 
quantities. This is done by exhibiting a generating function for such conserved quantities.

\subsection{Deriving integrable boundary conditions}\label{deriving_BC}

In order to start with an appropriate form for $P(x,t,k)$, we use the B\"acklund transformation method presented in \cite{H} and 
further developed in \cite{BT1,BT2} for the scalar nonlinear Schr\"odinger equation to treat a class of integrable boundary conditions. 
Contrary to the scalar case, in the vector case it is important to study both the $x$-part and the $t$-part of the equations defining 
the B\"acklund transformation to completely characterize integrable boundary conditions. This is done below. This is all the more important as 
the $t$-part of the B\"acklund method provides the link between this method and Fokas' approach in the case of linearizable boundary conditions.

We introduce a B\"acklund matrix $L(x,t,k)$ relating the auxiliary problem (\ref{equ:laxp1},\ref{equ:laxp2}) 
for $\Phi$ to the same auxiliary problem for $\widetilde{\Phi}$, with the potential $Q$ replaced by a new potential $\widetilde{Q}$, by the equation
\be
\label{2_copies}
\widetilde{\Phi}(x,t,k) = L(x,t,k) \Phi(x,t,k)\,.
\ee
It is well-known that $L$, also known as a gauge transformation of the auxiliary problem, satisfies the following equations
\bea
\label{equ:L1} 
  &&L_x + ik [ \Sigma _3 , L] = \widetilde{Q}\,L-LQ\,, \\
\label{equ:L2} 
  &&L_t + 2 i k^2 [\Sigma_3, L]=   \widetilde{G}\,L-LG\,,
\eea
and preserves the compatibility condition (or zero curvature condition). We look for a solution in the form
\be
L(x,t,k)=kI_{n+1}+A(x,t)\,,
\ee
under the symmetry constraint $\widetilde{Q}(x,t)=Q(-x,t)$. Writing $A$ in natural block form
\bea
\left(\begin{array}{cc}
A_1(x,t)&A_2(x,t)\\
A_3(x,t)&A_4(x,t)
\end{array}\right)\,,
\eea
and inserting in (\ref{equ:L1}) yields the equations
\begin{subequations}
\label{backlundrelations2}
\begin{align}
2 i A_2 (x,t) & =  R(-x,t) -  R(x,t)\,,\\
-2 i A_3 (x,t) & =  \lambda \left(R^\dagger (-x,t) -  R ^\dagger(x,t)\right) \,,
\end{align}
\end{subequations}
and
\begin{subequations}
\label{backlundrelations1}
\begin{align}
\label{backlundrelations1a} 
A_{1x} (x,t)& = R(-x,t) A_3(x,t) - \lambda A_2(x,t) R^\dagger (x,t)\,, \\  
\label{backlundrelations1b} 
A_{2x} (x,t)  & = R(-x,t) A_4(x,t) - A_1(x,t) R (x,t)\,, \\
\label{backlundrelations1c} 
A_{3x} (x,t)& =   \lambda \left[R^\dagger (-x,t) A_1(x,t) - A_4(x,t) R^\dagger (x,t)\right]\,, \\ 
\label{backlundrelations1d} 
A_{4x} (x,t) & =  \lambda R^\dagger (-x,t) A_2(x,t) - A_3(x,t) R (x,t)\,.
\end{align}
\end{subequations}
Note that 
\be\label{bound3}
A_3(x,t) =  \lambda A_2 ^\dagger(x,t)\,.
\ee
Combining (\ref{backlundrelations2}) and (\ref{backlundrelations1b},\ref{backlundrelations1c}) at $x=0$ yields the following boundary conditions
\begin{subequations}
\begin{align}
\label{bound4a}
R_x(0,t) = &- i (A_4(0,t)I_n - A_1(0,t))R(0,t)\,, \\
\label{bound4b}
R^\dagger_x(0,t) = & i R^\dagger(0,t)(A_1(0,t) - A_4(0,t)I_n) \,.
\end{align}
\end{subequations}
To ensure the compatibility between (\ref{bound4a}) and (\ref{bound4b}), we impose
\be \label{bound4}
A_1(0,t)-A_4(0,t)I_n= -(A_1(0,t)-A_4(0,t)I_n)^\dagger\,,
\ee 
showing that $H\equiv -i(A_1(0,t)-A_4(0,t)I_n)$ is a hermitian matrix.
The boundary condition now reads
\be
\label{general_Robin}
R_x(0,t)+ H R(0,t)=0\,. 
\ee
We note that at this stage, we have boundary conditions that depend on time a priori. We remove this time dependence by requiring that $A_1(0,t)$ and
$A_4(0,t)$ be time-independent. Then, H is time independent. We see that (\ref{general_Robin}) is the vector generalisation of the 
usual Robin boundary condition in the scalar case ($r_x(0,t)+\alpha r(0,t)=0$, $\alpha\in\RR$). The fact that $H$ is hermitian is the analog of $\alpha$
being real. Let us denote $A_4(0)=\beta$. What we have obtained so far reads
\bea
\label{L_x_part}
L(0,k)=kI_{n+1}+\left(\begin{array}{cc}
\beta I_n+iH & 0\\
0 & \beta
\end{array}\right)\,,
\eea
with $L$ independent of $t$ at $x=0$. To complete the characterization of $L(0,k)$, insert \eqref{L_x_part}
in (\ref{equ:L2}) to get
\bea
\widetilde{G}(0,k)\,L(0,k)-L(0,k)G(0,k)=0\,.
\eea
Noting that $\widetilde{G}(0,k)=\Sigma_3G(0,-k)\Sigma_3$, this reads
\bea
\label{link_Fokas}
G(0,-k)\,\Sigma_3 L(0,k)=\Sigma_3 L(0,k)G(0,k)\,.
\eea
It is now convenient to use the hermiticity of $H$ that guarantees that $H$ is diagonalisable by a unitary matrix $V$
\be
\label{diagonal}
H = V D V^\dagger, \qquad \text{where } D = \text{diag} \{ d_1, \dots, d_n \},
\ee
and the $d_j$'s are all real numbers. Note that the transformation $R(x,t)\mapsto V^\dagger R(x,t)=R^\prime(x,t)$ leaves the 
nonlinear Schr\"odinger equation invariant and 
in the new basis the boundary condition takes the simple, diagonal form
\be
\label{general_Robin_diag}
 R^\prime_x (0,t) + D  R^\prime(0,t)=0\,.
\ee
This shows that, in the presence of a boundary described by $H$, the VNLS equation has a preferred polarization basis determined by the boundary. 
In the following, we work in this basis and drop the $^\prime$. Then, 
\bea
L(0,k)=kI_{n+1}+\left(\begin{array}{cc}
\beta I_n+iD & 0\\
0 & \beta
\end{array}\right)\,,
\eea
and inserting the relation  $R_x(0,t)+ DR(0,t)=0$ in $G(0,k)$, equation (\ref{link_Fokas}) yields
\begin{subequations}
\begin{align}
(2i\beta I_n-D)DR(0,t)=0\,,\\  
R^\dagger(0,t)(2i\beta I_n-D)D=0\,, \\
RR^\dagger(0,t)D=DRR^\dagger(0,t)\,.
\end{align}
\end{subequations}
The compatibility of the first two equations imposes that $\beta$ be purely imaginary: $\beta=i\alpha$, $\alpha\in\RR$, unless $R(0,t)=0$ (this 
Dirichlet boundary condition for all the components is formally obtained when all the $d_j$ are infinite). Then, the first equation shows that 
either $d_j=0$ or $d_j=-2\alpha$ or $R_j(0,t)=0$. Finally, the last equation reads 
\be
d_j R_jR^*_k(0,t)=d_kR_jR^*_k(0,t)~~,~~j,k=1,\dots,n\,.
\ee
In general, this means that $d_j=d_k\equiv d$ \ie $D=dI_n$ is proportional to the identity matrix. The particular case $R_j(0,t)=0$ for some $j$ 
requires some attention. In this case, either $R_{jx}(0,t)$ is also zero and $d_j=d$ as before, or in general $R_{jx}(0,t)\neq 0$, meaning 
that $d_j=\infty$ is different from the common value $d$ and we must have $d_j+2\alpha=0$. So this case occurs when formally $\alpha=-\infty$. 
To summarize the results, we have the following possible integrable boundary conditions
\bea
R_x(0,t)-2\alpha R(0,t)=0~~,~~\alpha\in\RR\,,
\eea
or a mixture or Neumann and Dirichlet boundary conditions
\bea
&&R_j(0,t)=0~~,~~j\in S~~,\\
&&R_{kx}(0,t)=0~~,~~k\in \{1,\dots,n\}\setminus S\,,
\eea
where $S$ is a subset of $\{1,\dots,n\}$. In the rest of this paper, for convenience we will assume that $S$ is a proper subset of $\{1,\dots,n\}$. 
This avoids redundancy for the pure Dirichlet or Neumann condition which can be recovered from the Robin condition in the limits $\alpha\to\infty$ or 
$\alpha\to 0$ respectively.

In terms of $L(0,k)$ the previous results are more conveniently written by considering 
$$\cL(x,t,k)=\frac{1}{k+i\alpha}L(x,t,k)\,.$$ 
Note that $\cL$ is completely equivalent to $L$ since a B\"acklund matrix is always defined up to a function of $k$ but it has the advantage of 
accommodating the $\alpha=-\infty$ case. Then, the previous two cases correspond to 
\bea
\cL(0,k)=\left(\begin{array}{cc}
\frac{k-i\alpha}{k+i\alpha}I_n & \\
 & 1
\end{array}\right)~~\text{or}~~\cL(0,k)=\left(\begin{array}{cccc}
\sigma_1& & & \\
&\ddots& & \\
& & \sigma_n &\\
& & & 1
\end{array}\right)\,,
\eea
where $\sigma_j=-1$, $j\in S$ and $\sigma_j=1$, $j\in  \{1,\dots,n\}\setminus S$.

We note that in \cite{HS}, the previous mixed Neumann/Dirichlet boundary condition was not found. In section \ref{examples}, we will see that 
this is the most interesting one as far as interactions between vector solitons and boundary are concerned.

It is now worthwhile to comment on relation (\ref{link_Fokas}). This is precisely the relation that is imposed in the Fokas' method to obtain 
the so-called linearizable boundary conditions (with the identification $\Sigma_3L(0,k)\equiv N(k)$, see e.g. \cite{fokasbook}). Thus, for each 
solution $N(k)$ of equation (\ref{link_Fokas}) that can be identified as the value at $x=0$ of a B\"acklund matrix $L(x,t,k)$, we see that 
the so-called linearizable boundary conditions are integrable. In fact this is shown in section \ref{integrability} where a generating 
function for the conserved quantities is constructed for the class of boundary conditions described by such an $N(k)$.

\subsection{Integrability}\label{integrability}

The fact that our boundary conditions derive from a B\"acklund matrix ensures the existence of an infinite number of conserved quantities. 
This is what we mean by integrability in this context. Adapting a
well-known argument for deriving a generating function of conserved quantities from the Lax pair to our situation, we show this explicitely below. 
Note that we expect integrability in the Liouville sense to hold as well once one equips the present system with a Hamiltonian structure. Indeed, 
B\"acklund transformations are known to have nice canonical properties (see e.g. \cite{Skly_cano}).
\begin{proposition}
\label{prop_generating}
A generating function for the conserved quantities of the VNLS with integrable boundary conditions is given by
\bea
\label{symmetrized}
i(I(k)-I^\dagger(k^*))\,,
\eea
where
\bea
I(k)&=&\text{tr}\left[\int_0^\infty \,\left[R(x,t)(\Gamma(x,t,k)-\Gamma(x,t,-k))\right]\,dx\right]\,,
\eea
and $\Gamma(x,t,k)$ satisfies the following Ricatti equation
\bea
\label{Ricatti}
\Gamma_x=2ik\Gamma+ \lambda R^\dagger-\Gamma R\Gamma\,.
\eea
\end{proposition}
\prf The proof follows the one presented in \cite{VC} which we adapt to the present vector case with a boundary. 
Recall that we consider two copies of the auxiliary problem (\ref{equ:laxp1},\ref{equ:laxp2}) related by 
(\ref{2_copies}). All the matrices involved in the argument, like $\Phi$, $G$ or $L$ are split in natural blocks
\bea
Z=\left(\begin{array}{cc}
Z_{11}&Z_{12}\\
Z_{21}&Z_{22}
\end{array}\right)~~,~~Z=\Phi,G,L\,,
\eea
where $Z_{11}$ is an $n\times n$ matrix, $Z_{12}$ is an $n$-component column vector, etc.
Define $\Gamma(x,t,k)=\Phi_{21}\Phi_{11}^{-1}(x,t,k)$ and 
$\widetilde{\Gamma}(x,t,k)=\widetilde{\Phi}_{21}\widetilde{\Phi}_{11}^{-1}(x,t,k)$. Then, (\ref{equ:laxp1}) yields 
(\ref{Ricatti}) for $\Gamma$ and the same equation for $\widetilde{\Gamma}$ with $R$ replaced by $\widetilde{R}$. 
Also, we can write 
\bea
\Phi_{11x}\Phi_{11}^{-1}=R\Gamma~~,~~\Phi_{11t}\Phi_{11}^{-1}=G_{11}+G_{12}\Gamma\,,
\eea
so using $(\ln \Phi)_{xt}=(\ln\Phi)_{tx}$ we get $(R\Gamma)_t=(G_{11}+G_{12}\Gamma)_x$ where $G_{ij}$ are the 
appropriate blocks of $G$. A similar relation holds for $\widetilde{\Gamma}$. Finally, from (\ref{2_copies}) we have
\bea
\widetilde{\Phi}_{11t}&=&\left[(L_{11}+L_{12}\Gamma)\Phi_{11}\right]_t\nonumber\\
&=&\left[(L_{11}+L_{12}\Gamma)_t+(L_{11}+L_{12}\Gamma)(G_{11}+G_{12}\Gamma)\right](L_{11}+L_{12}\Gamma)^{-1}\widetilde{\Phi}_{11}\,,
\eea
which we compare to $\widetilde{\Phi}_{11t}=(\widetilde{G}_{11}+\widetilde{G}_{12}\widetilde{\Gamma})\widetilde{\Phi}_{11}$ to get 
\bea
\text{tr}(\widetilde{G}_{11}+\widetilde{G}_{12}\widetilde{\Gamma})=\text{tr}(G_{11}+G_{12}\Gamma)+\text{tr}\ln(L_{11}+L_{12}\Gamma)_t\,.
\eea
Swapping the roles of $\Phi$ and $\widetilde{\Phi}$ and introducing $\widetilde{L}=L^{-1}$ we can also obtain
\bea
\text{tr}(G_{11}+G_{12}\Gamma)=\text{tr}(\widetilde{G}_{11}+\widetilde{G}_{12}\widetilde{\Gamma})+\text{tr}\ln(\widetilde{L}_{11}+\widetilde{L}_{12}\widetilde{\Gamma})_t\,.
\eea
This allows us to obtain a more symmetric form of the final result. Putting everything together, we get the general result
\bea
\label{general_conservation}
\partial_t\text{tr}\left[\int_{-\infty}^0 \widetilde{R}\widetilde{\Gamma}(x,t,k)\,dx+\int_0^\infty R\Gamma(x,t,k)\,dx \right]\nonumber 
\qquad\qquad\qquad\qquad\qquad\qquad\qquad\quad\\
=
\frac{1}{2}\partial_t\text{tr}\left[ \ln \left(L_{11}(0,k)+L_{12}(0,k)\Gamma(0,t,k)\right)-\ln \left(\widetilde{L}_{11}(0,k)+\widetilde{L}_{12}(0,k)
\widetilde{\Gamma}(0,t,k)\right) \right].
\eea
Now, for the problem with boundary, the thing to note is that under the reduction $\widetilde{R}(x,t)=R(-x,t)$, we have 
\bea
\widetilde{\Gamma}(x,t,k)=-\Gamma(-x,t,-k)~~\text{and}~~\widetilde{L}(x,t,k)=\frac{1}{k^2+\alpha^2}\,\Sigma_3 L(-x,t,-k)\Sigma_3\,.
\eea
Finally, at $x=0$, $L(0,k)$ does not depend on $t$ for the class of boundary conditions we have derived and $L_{12}(0,k)=0$ so the right-hand-side in 
(\ref{general_conservation}) vanishes. Therefore, we have shown 
\bea
\partial_t I(k)=0\,.
\eea
The special form (\ref{symmetrized}) is used to get real conserved quantities.
\finprf

In practice, the conserved quantities are determined recursively by inserting the following expansion
\bea
\Gamma(x,t,k)=\sum_{n=1}^\infty \frac{\Gamma_n(x,t)}{(2ik)^n}\,,
\eea
into the Ricatti equation to obtain
\bea
\Gamma_1=-\lambda R^\dagger~~,~~\Gamma_{n+1}=\Gamma_{nx}+\sum_{k=1}^{n-1}\Gamma_k R \Gamma_{n-k}~~,~~n\ge 1\,.
\eea
As expected from the presence of the boundary, we see that the conserved quantities corresponding to even powers of $1/k$ do not appear. In particular, 
the momentum (order $1/k^2$) is not conserved.

The connection between linearizable boundary conditions and B\"acklund transformations was noted in \cite{Fokas_sG}. Here, in view of the 
previous result, this connection allows one to identify explicitely the infinite set of conserved quantities once a solution $\Sigma_3 L(0,k)\equiv N(k)$ of
 the "symmetry" relation \eqref{link_Fokas} is known. Of course, this connection goes beyond the specific example of VNLS and holds for any integrable system characterized by a Lax pair 
(with possible complications for equations that are not invariant under $x\to -x$ such as the KdV equation).

\subsection{Mirror image construction}\label{mirror_image}

Following \cite{Fokas2,BH}, we can now use the result of the previous section to introduce the following extended potential 
\bea
P(x,t,k)=\theta(x)Q(x,t)+\theta(-x)\cB(k)Q(-x,t)\cB(-k)\,,
\eea
where $\theta$ is the Heaviside function and 
\bea
\label{form_B}
\cB(k)\equiv\bma
B(k)&0\\
0&-1
\ema=\Sigma_3\,\cL(0,k)\,.
\eea
Note that
\bea
\label{properties_B}
B^\dagger(k^*)B(k)=I_n=B(-k)B(k)~,~~k\in\CC\,.
\eea
We have the property
\bea
\label{symmetry_P}
P(-x,t,-k)=-\cB(-k)P(x,t,k)\cB(k)\,.
\eea
The idea is to apply the inverse scattering method as explained in section \ref{sec:2} to $P$ instead of $Q$. 
By construction, for $x>0$, any solution found for $P$ gives a solution $Q$ of VNLS on the half-line. The delicate point is to check 
whether this solution does indeed satisfy the boundary conditions that we derived in the previous section and that we encoded in Fourier space in the matrix 
$B(k)$ obtained from $\cL(0,k)$.

For conciseness, we keep the same notations as in the previous section but with $Q$ replaced by $P$. Using (\ref{symmetry_P}), we obtain
that $M_+(x,t,k)$ and $\cB(k)M_-(-x,t,-k)\cB(-k)$ satisfy the same Volterra integral equation. Noting that they have the same asymptotics as $x\to\infty$ we deduce
\bea
\label{symmetry_relation}
M_+(x,t,k)=\cB(k)M_-(-x,t,-k)\cB(-k)\,.
\eea
From this and (\ref{def_S}), we obtain
\bea
S^{-1}(k)=\cB(k)S(-k)\cB(-k)\,.
\eea
Next, from (\ref{equ:inv}), we have that $WM^\dagger_\pm(x,t,k^*)W^{-1}=M_\pm^{-1}(x,t,k)$ which in turn yields\footnote{We note that strictly speaking, 
$S(k)$ is defined first for $k\in\RR$. But as we have seen, its entries can be analytically continued to various parts of the complex plane. Eq 
(\ref{dag_symmetry_S}) is then to be understood as relations for the entries of $S$ valid wherever they make sense. Consequently, the same is true 
of eq (\ref{mirror_dag_symmetry}).}
\bea
\label{dag_symmetry_S}
S^{-1}(k)=WS^\dagger(k^*)W^{-1}\,.
\eea
Combining the two previous equations yields the important result
\bea
\label{mirror_dag_symmetry}
WS^\dagger(k^*)W^{-1}=\cB(k)S(-k)\cB(-k)\,.
\eea
In components, this reads
\bea
\label{symmetries_scattering_data1}
a^\dagger(k^*)=B(k)a(-k)B(-k)~,~~\bar{a}^*(k^*)=\bar{a}(-k)\,,\\
\label{symmetries_scattering_data2}
\bar{b}^\dagger(k^*)=\lambda b(-k)B(-k)~,~~\lambda b^\dagger(k^*)=B(k)\bar{b}(-k)\,.
\eea
Note that due to the property $B^\dagger(k^*)=B(-k)$, the last two relations are consistent. The first relation implies that if $k_j$ is 
a zero of $\det a(k)$ then $-k_j^*$ is also a zero. The same holds true for $\bar{a}(k)$. Therefore, we find that the main observation of \cite{BH} that 
the relevant zeros to formulate the inverse problem come in quartets $\{\pm k_j,\pm k_j^*\}$ is also valid in the vector case. The total number of zeros 
in each half-plane is even: $J=2N$ and there are $N$ zeros in each quadrant of the complex plane. For each $j=1,\dots,N$, it is enough to assume that 
$k_j$ is in the first quadrant \ie $\text{Re}\, k_j\ge 0$ and $\text{Im}\, k_j> 0$. The other three zeros are obtained by successive reflections with 
respect to the imaginary and real axes.

Associated to these zeros are norming constants which we denote $C_j$ (for $k_j$) and $C_j{'}$ (for $-k_j^*$). From the symmetry 
(\ref{equ:inv}), the norming constants associated to $k_j^*$ (resp. $-k_j$) are then $\lambda C_j^\dagger$ 
(resp. $\lambda C_j^{'\dagger}$).

The subtle point of the method now is to reconstruct $R(x,t)$ from the previous scattering data corresponding to $P(x,t,k)$. This is achieved by 
establishing the analogs of (\ref{coupled_eq1}), (\ref{coupled_eq2}) and (\ref{R_reconstructed}) and restricting to $x>0$ where $P(x,t,k)=Q(x,t)$.
To do so we must investigate the analytic properties of $M_\pm(x,t,k)$ as functions of $k$ for $x>0$. The argument follows that given in \cite{Fokas2}.
For $x>0$, $P(x,t,k)=Q(x,t)$ so the analytic properties of $M_+$ are the same as in the infinite line case. Then, the analytic properties of $M_-$ 
are obtained using the symmetry relation (\ref{symmetry_relation}). In the mixed Neumann/Dirichlet case, no extra pole arises from the presence of $B(k)$
(which is in fact independent of $k$). In the Robin case, $M$ gets an extra pole at $k=i\alpha$ if $\alpha>0$ and $\overline{M}$ gets an extra pole 
at $k=-i\alpha$ if $\alpha>0$ but are otherwise analytic in their respective half-plane. Then, $a(k)$ and $\bar{a}(k)$ share the 
same analytic properties as $M$ and $\overline{M}$. As a consequence, the fundamental quantities $Ma^{-1}$ and $\overline{M}\bar{a}^{-1}$ have at worst a 
removable singularity. Therefore, the analysis of section \ref{ISP} goes through as before, taking into account 
the additional symmetry properties of the scattering data discussed above.
This is summarized in the following proposition which constitutes the main result of this work.
\begin{proposition}
\label{main_result}
The reconstruction formulae for $Q(x,t)$, $x>0$ are given by (\ref{coupled_eq1}), (\ref{coupled_eq2}) and (\ref{R_reconstructed}) with $J=2N$ and, 
without loss of generality, the substitution $k_{2j-1}\to k_j$, $j=1,\dots,N$ with corresponding norming constants $C_j$ and $k_{2j}\to -k_j^*$, $j=1,\dots,N$ 
with corresponding norming constants $C_j^{'}$. The scattering data involved in the formula satisfy the symmetry relations (\ref{symmetries_scattering_data1},
\ref{symmetries_scattering_data2}) and
\bea
\label{relations_norming_constants}
\lambda C_j^{'\dagger}C_j=B(-k_j)\frac{A(k_j)^t}{\alpha^{'2}(k_j)}~~,~~j=1,\dots,N\,.
\eea
\end{proposition}
\prf
The only thing left to prove are the relations (\ref{relations_norming_constants}) between the norming constants $C_j$ of the real solitons (those living 
on $x>0$), and the norming constants $C_j^{'}$ of the mirror solitons (those living on $x<0$). We use the following chain of relations obtained from (\ref{def_S}) and (\ref{symmetry_relation}) evaluated at $k=k_j$ and 
$k=-k_j$ 
\bea
M(x,t,k_j)A^t(k_j)&=&e^{2i\phi(x,t,k_j)}N(x,t,k_j)b(k_j)A^t(k_j)\,,\\
N(x,t,k_j)&=&\left(\begin{array}{cc}
-B(k_j)& \\
 & 1
\end{array}\right)\overline{M}(-x,t,-k_j)\,,\\
\overline{M}(-x,t,-k_j)&=&e^{-2i\phi(-x,t,-k_j)}\overline{N}(-x,t,-k_j)\bar{b}(-k_j)\,,\\
\overline{N}(-x,t,-k_j)&=&\left(\begin{array}{cc}
B(-k_j)& \\
 & -1
\end{array}\right)M(x,t,k_j)B(k_j)\,.
\eea
The compatibility of these relations yields 
$A^t(k_j)=-B(k_j)\bar{b}(-k_j)b(k_j)A^t(k_j)$ which is just (\ref{relations_norming_constants}) upon using $C_j=\frac{1}{\alpha^{'}(k_j)}b(k_j)A(k_j)^t$, 
$\bar{C}_j^{'}=\frac{\bar{b}(-k_j)}{\bar{a}^{'}(-k_j)}=\lambda C_j^{'\dagger}$ and $\bar{a}^{'}(-k_j)=-\alpha^{'}(k_j)$.
\finprf
Note that following the same argument but for $k=-k_j^*$ and $k=k_j^*$, we arrive at the following similar relations
\bea
\label{relations_norming_constants2}
\lambda C_j^{\dagger}C_j^{'}=B(k_j^*)\frac{A(-k_j^*)^t}{\alpha^{'2}(-k_j^*)}~~,~~j=1,\dots,N\,,
\eea
which are compatible with (\ref{relations_norming_constants}).

The symmetries of the scattering data ensure that $Q$ satisfies the desired boundary conditions. We show this in general in the mixed 
Neumann/Dirichlet case.

\begin{proposition}
\label{mixed_symmetry}
Without loss of generality take $B(k)=diag(1,\dots,1,-1,\dots,-1)\equiv B$ then the reconstruction formulae yield a potential $R$ which satisfies 
\bea
R(-x,t)=BR(x,t)\,,
\eea
\ie $R$ satisfies mixed Neumann/Dirichlet boundary conditions.

\end{proposition}
\prf
In the case of fast decreasing solutions, the reconstruction formula (\ref{R_reconstructed}) can be compactly rewritten as 
\bea
\label{formula_R1}
R(x,t)=\frac{1}{\pi}\int_{\overline{C}} e^{-2i\omega x-4i\omega^2t}\overline{N}^{(up)}(x,t,\omega)\bar{\rho}(\omega)\,d\omega\,,
\eea
where $\overline{C}$ is a contour in the lower half-plane from $-\infty$ to $+\infty$ that passes below all the zeros of $\bar{a}(k)$ and that is 
symmetric with respect to the imaginary axis. Repeating the arguments of section \ref{ISP} from the relation
\bea
e^{i\phi(x,t)\Sigma_3} M_-(x,t,k) e^{-i\phi(x,t)\Sigma_3}\, S^{-1}(k)=e^{i\phi(x,t)\Sigma_3} M_+(x,t,k) e^{-i\phi(x,t)\Sigma_3}\,,
\eea
instead of (\ref{def_S}), one can derive an equivalent reconstruction formula
\bea
R(x,t)=-\frac{1}{\pi}\int_{C} e^{-2i\omega x-4i\omega^2t}M^{(up)}(x,t,\omega)\tau(\omega)\,d\omega\,,
\eea
where $M$ (and $\overline{M}$) satisfy coupled integral equations similar to (\ref{coupled_eq1},\ref{coupled_eq2}) and $\tau(k)=d(k)c^{-1}(k)$ is 
obtained from the entries of $S^{-1}(k)$
\bea
S^{-1}(k) = \left(\begin{array}{cc}
\bar{c}(k)&d(k)\\
\bar{d}(k)&c(k)
\end{array}\right)\,.
\eea
Now combining (\ref{dag_symmetry_S}) and (\ref{mirror_dag_symmetry}), we get $\tau(k)=-B\bar{\rho}(-k)$. Also, from (\ref{symmetry_relation}), we 
get $\overline{N}^{(up)}(-x,t,-k)=BM^{(up)}(x,t,k)B$. Finally, performing the change of variable $\omega\to -\omega$ in (\ref{formula_R1}) and 
using the previous relations yields the result.
\finprf

To the authors' knowledge, a general proof that the field obtained with the mirror image technique obeys the desired boundary conditions
is not known within this formalism. Even in the scalar case, an explicit argument has been given in \cite{Fokas2} 
only in the case of Dirichlet boundary conditions. Arguments based on the linear limit of the nonlinear Schr\"odinger equation in the case 
of general boundary conditions have been given in \cite{BH}. It seems that the proper way to attack the problem is to use the general method developed 
by Fokas and co-workers in the case of linearizable boundary conditions. In view of the connection that we have made
above between this method and the B\"acklund transformation approach that we have used here in the vector case, 
we see that a full treatment of the problem requires to formulate the vector version of 
Fokas method. Hopefully, this problem is still amenable for the class of linearizable boundary conditions that we consider here. 
This is beyond the scope of the present paper and is left for future investigation.

It is worth noting the appearance of the matrix $A^t(k_j)$ in the relations between the norming constants of the "real" solitons and 
those of the mirror image solitons. This is a new feature compared to the scalar case. Using the dressing method \cite{ZS1}, it can be shown that 
these matrices can be expressed in terms of all the norming constants. Hence, the system (\ref{relations_norming_constants}) is a system 
of coupled matrix equations that one has to solve in order the completely characterize the soliton solutions with boundary. This is technically 
more challenging than in the scalar case where one could use directly the trace formula for $a(k)$ to solve the 
problem \cite{BH}. This is reminiscent of the difference between the scalar NLS and the Manakov model on the line where the presence of 
polarization vectors for the solitons greatly complicates the analysis of the soliton interactions \cite{Tsu}. 
Here, the presence of the cofactor matrix captures the influence 
of the polarization vectors of the solitons on their interaction with the boundary. 

A general efficient algorithm has not yet been found to solve these coupled equations and completely characterize the $N$-soliton solutions 
with a boundary. We hope to return to this question in a future work. In the next section, we show how this can be done for one soliton and use the 
result to investigate its interaction with the boundary.

\section{One soliton reflection: transmission between modes}\label{examples}

We consider the pure soliton case with $N=1$ \ie we study the behaviour of one soliton bouncing off the boundary. The scattering data is $k_1$ 
and $C_1$. From this, we deduce the scattering data corresponding to the mirror soliton: $-k_1^{*}$ and $C_1^{'}$ 
determined by (\ref{relations_norming_constants2}) (with $\lambda=-1$)
\bea
\label{constraint}
C_1^{\dagger}C_1^{'}=-B(k_1^*)\frac{A(-k_1^*)^t}{\alpha^{'2}(-k_1^*)}\,.
\eea

We use the dressing method to compute $A(-k_1^*)^t$. Let us first recall this method to generate two arbitrary solitons from the 
vacuum solution and then we will apply it in the present special case. Let $\kappa_1$, $\kappa_2$, ${\cal C}_1$ and ${\cal C}_2$ be the poles and norming 
constants for the two solitons. The scattering coefficient $a(k)$ is then given by 
\bea
a(k)=D_1(k)D_2(k)\,,
\eea
where 
\bea
D_j(k)=I_n+\left(\frac{k-\kappa_j}{k-\kappa_j^*}-1\right)\Pi_j\,,
\eea
and $\Pi_j$ is a (one-dimensional) orthogonal projector obtained recursively by requiring ${\cal C}_j\,a(\kappa_j)=0$, $j=1,2$. From this, one easily deduces 
\bea
\frac{A(\kappa_2)^t}{\alpha^{'}(\kappa_2)}=(\kappa_2-\kappa_2^*)\Pi_2D_1(\kappa_2)^{-1}\,,
\eea
with ${\cal C}_1\Pi_1={\cal C}_1$ and $\rho_2\Pi_2=\rho_2$ where $\rho_2={\cal C}_2D_1(\kappa_2)$.

Now, we insert this in (\ref{constraint}) with the identification $\kappa_1=k_1$, $\kappa_2=-k_1^*$, ${\cal C}_1=C_1$ and 
${\cal C}_2=C_1^{'}$. Multiply on the right by $D_1(\kappa_2)$ and project on $\rho_2^\dagger$ to obtain
\bea
\frac{\rho_2^\dagger}{\rho_2\rho_2^\dagger}=-\frac{\alpha^{'}(-k_1^*)}{k_1-k_1^*}B(-k_1^*)C_1^\dagger\equiv V_1^\dagger\,.
\eea
Hence, we deduce the solution
\bea
C_1^{'}=\frac{V_1}{V_1V_1^\dagger}D_1(-k_1^*)^{-1}\,.
\eea
For practical calculations, we note finally that $\alpha^{'}(-k_1^*)$ can be obtained from the well-known trace formula for $\det a(k)$ (see e.g. \cite{APT})
which gives, in our case,
\bea
\alpha^{'}(-k_1^*)=\frac{k_1+k_1^*}{2k_1^*(k_1-k_1^*)}\,.
\eea

In the following, we present numerical results for the case of two components ($n=2$) and for the mixed Neumann/Dirichlet case. The Robin boundary condition
does not bring anything new in the vector case as compared to the scalar case in the sense that in any polarization basis, each component of $R$ 
always satisfies a scalar Robin boundary condition. However, in the mixed Neumann/Dirichlet case, if the polarization basis is different from the preferred 
boundary basis \ie if we restore the unitary $V$ matrix in (\ref{diagonal}), then an interesting phenomenon of reflection-transmission between the modes 
appears: the amplitude of the soliton envelope $|R_j(x,t)|$ is different before and after its interaction with the boundary!

As an example, we take 
\bea
V=\left(\begin{array}{cc}
\cos \theta e^{i\zeta}&\sin\theta e^{i\xi}\\
-\sin\theta e^{-i\xi}&\cos\theta e^{-i\zeta}
\end{array}\right)\,,
\eea
with $\theta,\xi,\zeta\in\RR$ and consider the boundary matrix $B_\theta$ given by
\bea
B_\theta=V\left(\begin{array}{cc}
1&0\\
0&-1
\end{array}\right)V^{-1}\,.
\eea
The parameters $\theta,\zeta,\xi$ measure the "deviation" from the natural boundary basis corresponding to $\theta=\zeta=\xi=0$.
Below are plots of $|R_1|$ and $|R_2|$ in $x,t$ space for $k_1=1+\frac{i}{2}$ and $C_1=(2~~1)^t$.\\ 

\begin{figure}[htp]
\begin{minipage}[t]{8cm}
\epsfig{file=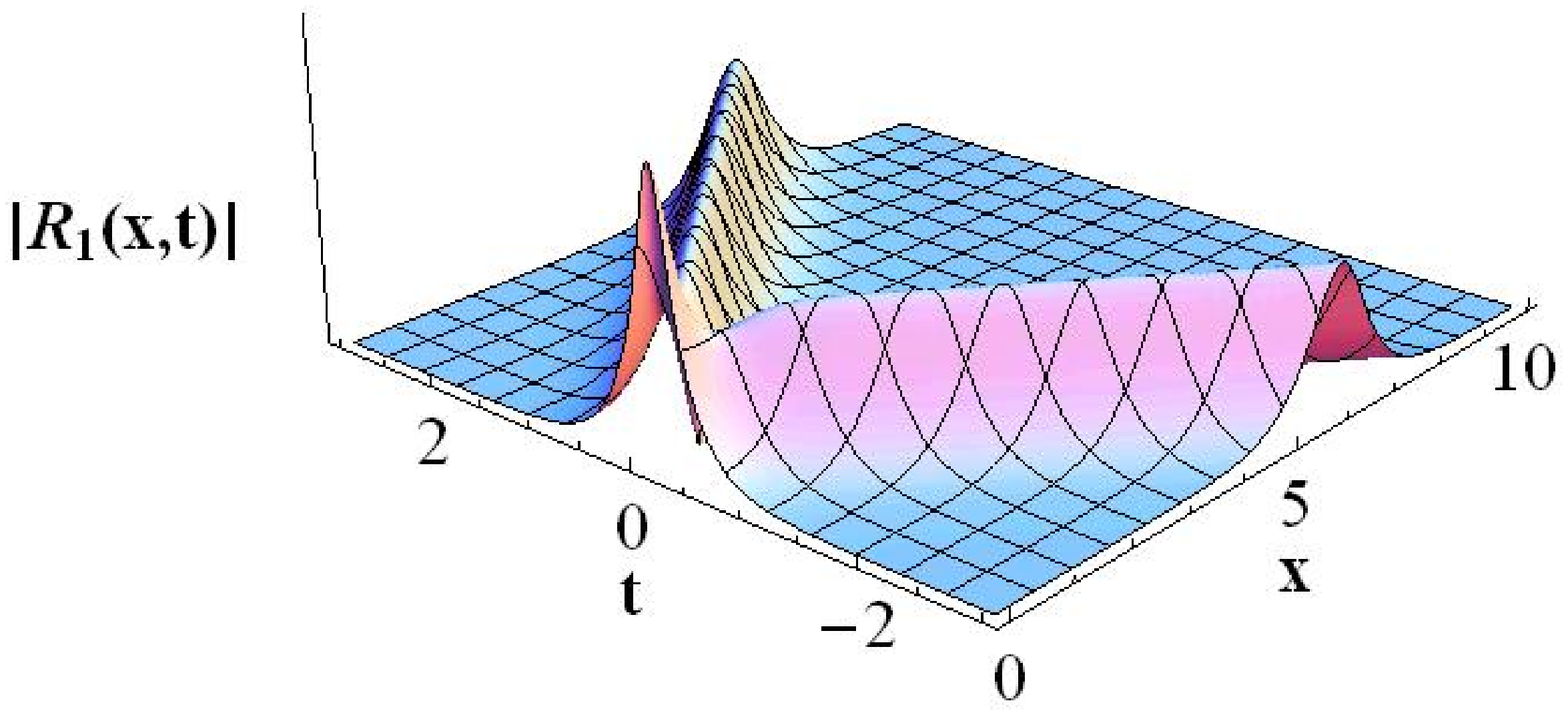,width=8cm}
\caption{\label{fig-R1-N} Reflection of the first component of the soliton on the boundary with Neumann boundary condition. 
The amplitude before and after the reflection is the same.}
\end{minipage}
\qquad
\begin{minipage}[t]{8cm}
\epsfig{file=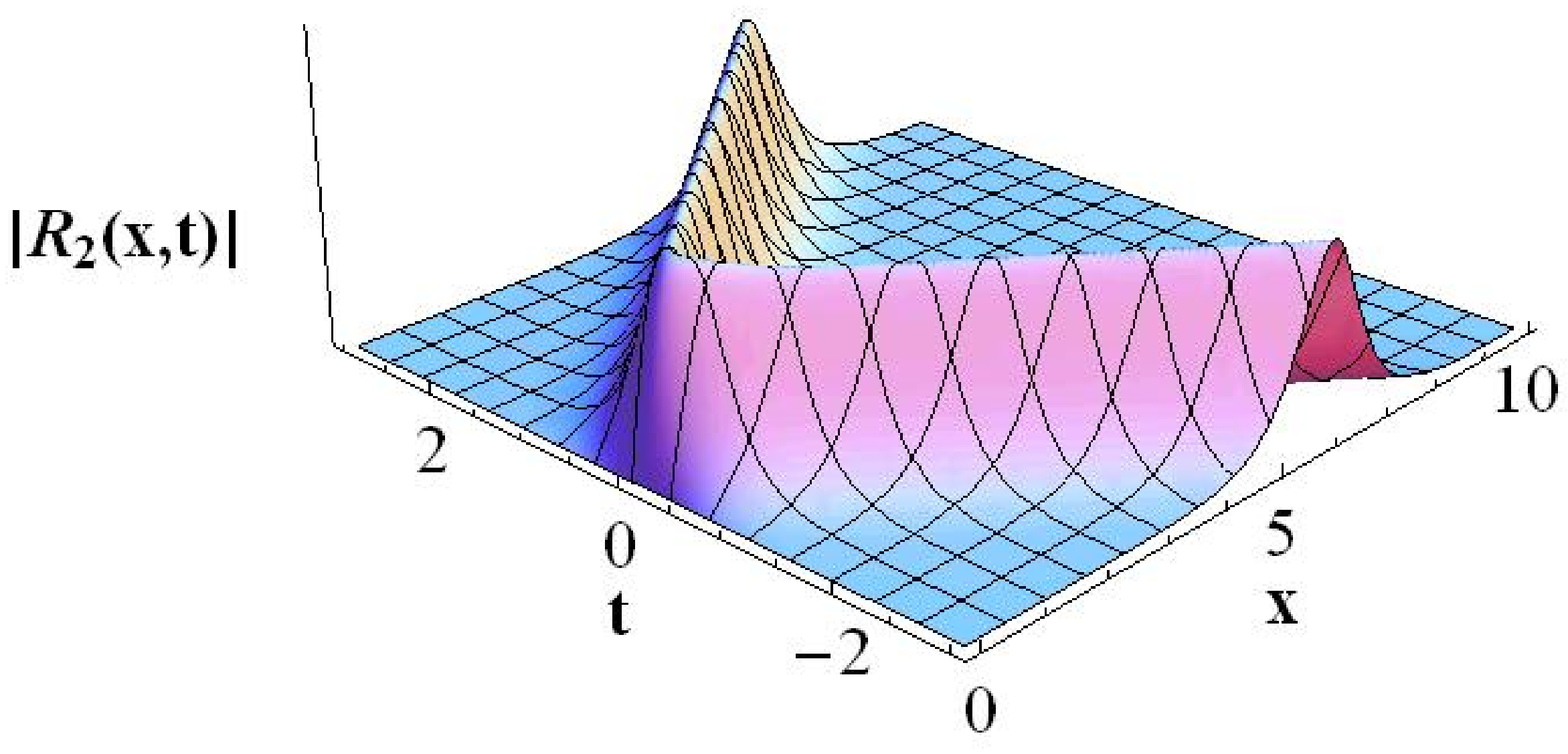,width=8cm}
\caption{\label{fig-R2-D} Reflection of the second component of the soliton on the boundary with Dirichlet boundary condition. 
The amplitude before and after the reflection is the same.}
\end{minipage}
\end{figure}

First, for $\theta=\zeta=\xi=0$ (Fig. \ref{fig-R1-N} and \ref{fig-R2-D}) we see that, as expected, the first mode satisfies a Neumann boundary condition while the second mode satisfies a Dirichlet boundary 
condition. These plots are very similar to those of \cite{BH} in each case. Each mode behaves like a scalar solution seeing its own boundary condition.

\begin{figure}[htp]
\begin{minipage}[t]{8cm}
\epsfig{file=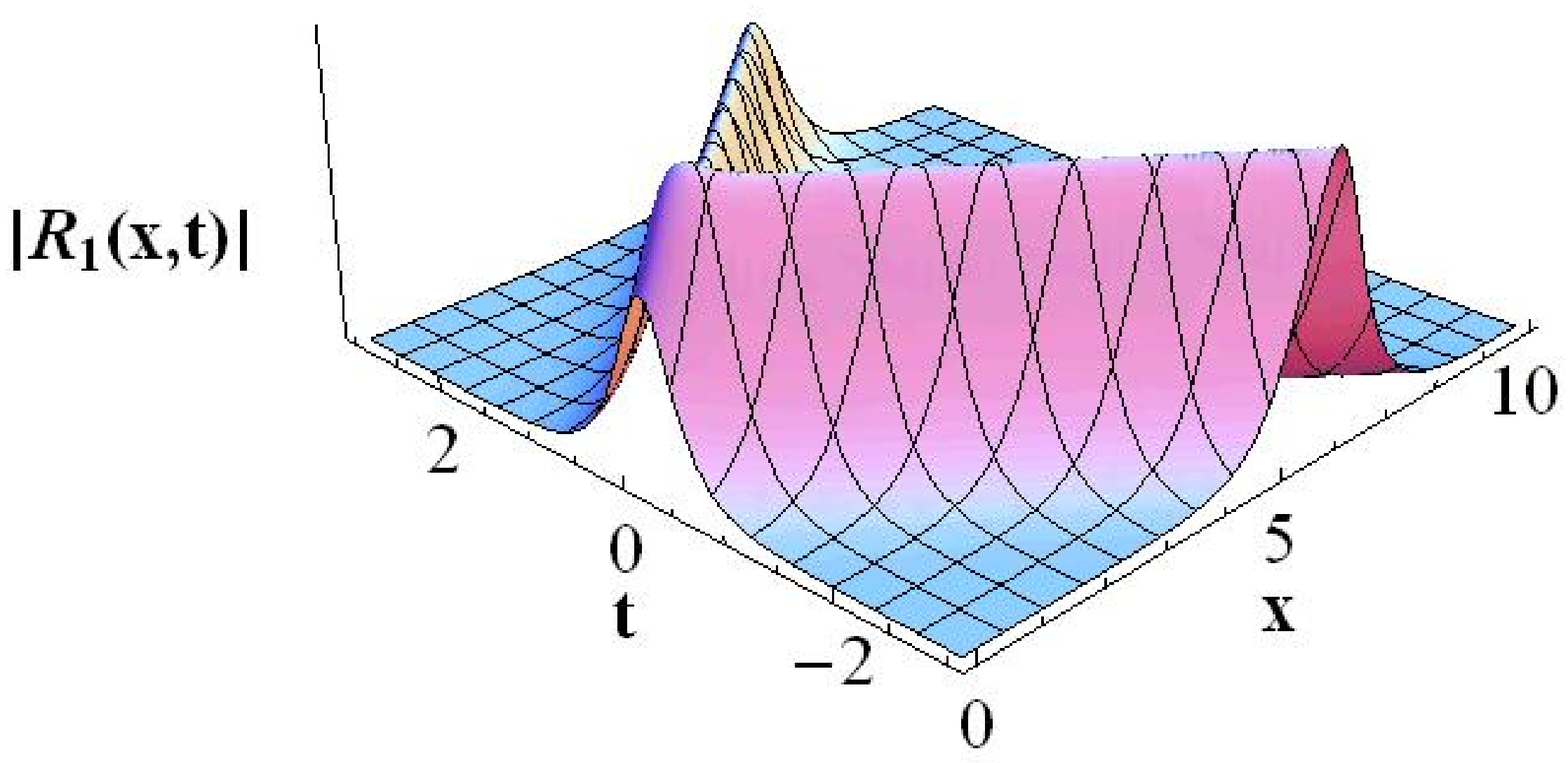,width=8cm}
\caption{\label{fig-R1}Reflection of the first component of the soliton on the boundary. 
The outgoing mode has a smaller amplitude than the incoming one.}
\end{minipage}
\qquad
\begin{minipage}[t]{8cm}
\epsfig{file=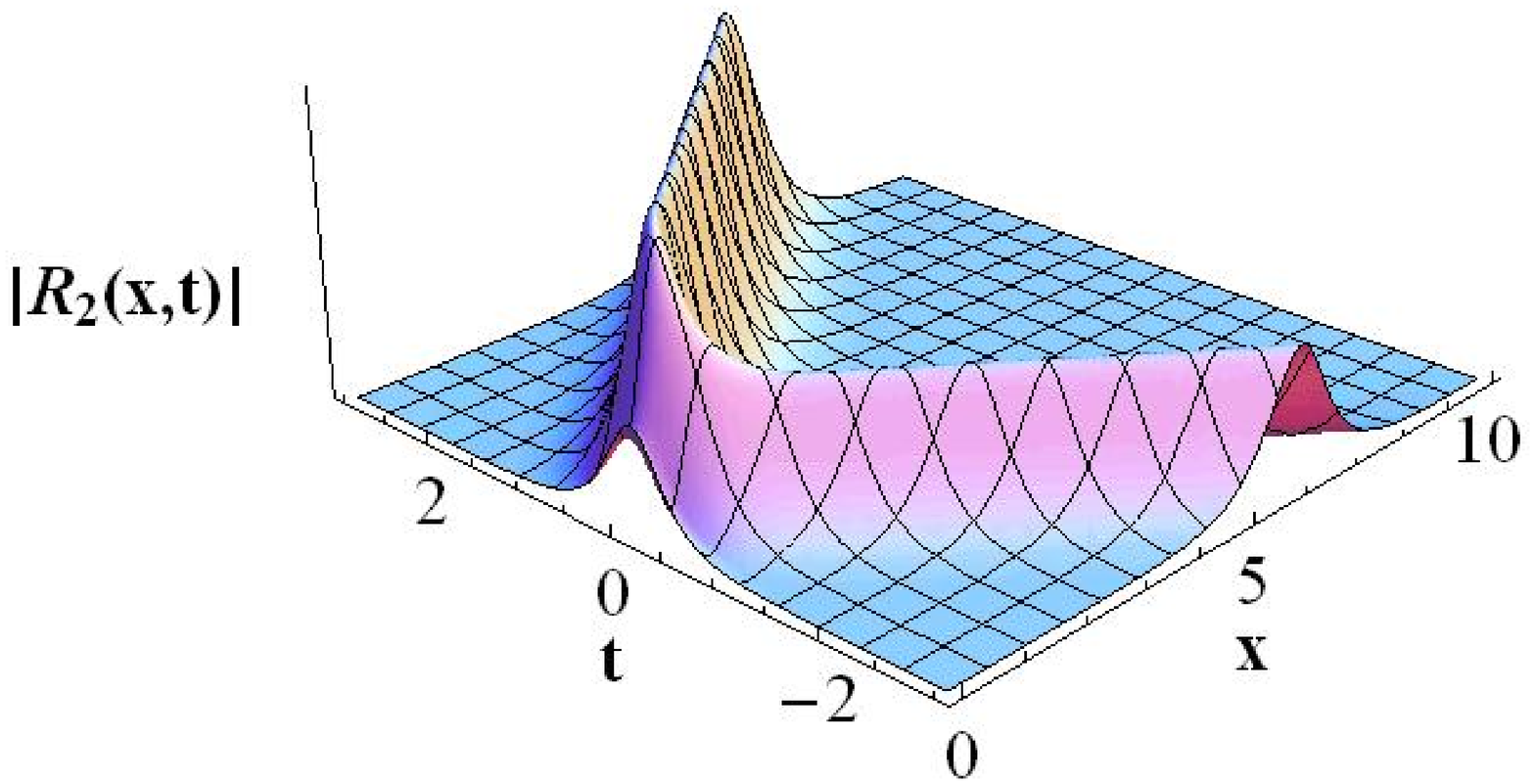,width=8cm}
\caption{\label{fig-R2} Reflection of the second component of the soliton on the boundary. Part of the first mode 
has been transmitted to the outgoing mode here.}
\end{minipage}
\end{figure}

Now, for $\theta=\frac{\pi}{6}$ and $\zeta=\xi=0$ (Fig. \ref{fig-R1} and \ref{fig-R2}), we clearly see that the 
amplitude of each mode before and after the interaction with the boundary is different. Both modes are reflected but part of mode $1$ is 
transmitted to mode $2$.

This shows that the boundary acts as some sort of polarization filter. We emphasize though that there is no loss in the transmission process in the sense 
that the quantity $|R_1|^2+|R_2|^2$ is indeed a conserved density. Another way to see this is to investigate the same quantity in the reference frame 
of the incoming or outgoing soliton. Here the velocity of the incoming (outgoing) soliton is $-v$ ($v$) with $v=4\,\text{Re}\,k_1=2$. We find indeed that 
\bea
\lim_{t\to-\infty}(|R_1|^2+|R_2|^2)(-vt+x_{in},t)=\lim_{t\to\infty}(|R_1|^2+|R_2|^2)(vt+x_{out},t)\,,
\eea
where $x_{in}$ ($x_{out}$) is the incoming (outgoing) soliton centre.

One can then wonder if one can use the boundary to control the amplitudes of the modes by changing the values of the boundary parameters. 
The following plot shows that this can be done. We keep $k_1=1+\frac{i}{2}$ and $C_1=(2~~1)^t$ and we vary $\theta$ from $0$ to $\frac{\pi}{2}$ while 
fixing $\zeta=1.11$ and $\xi=0$. 
The incoming amplitudes are of course constant (lines) while the outgoing amplitudes vary (dashed curves). In our example, we see that the 
second mode (in blue) can be made vanishingly small (for $\theta\approx 1.15$). The total outgoing amplitude is constant though (black line) and equals
the total incoming amplitude.
\begin{figure}[htp]
\begin{center}
\epsfig{file=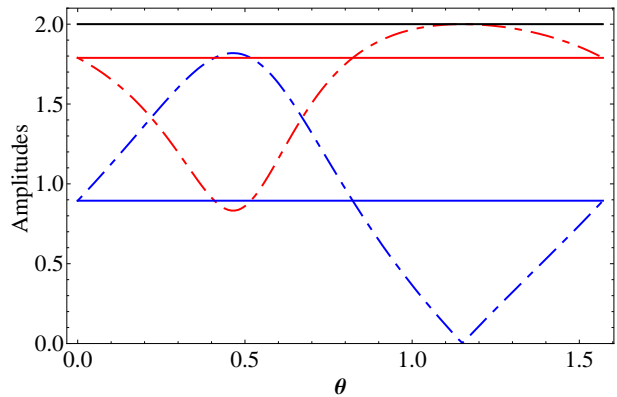,width=8cm}
\caption{\label{fig-amp}Amplitudes in the soliton frame as functions of $\theta$. The red line is the incoming $|R_1|$ amplitude. The blue line is incoming $|R_2|$. 
Dashed curves are the corresponding outgoing amplitudes. The black line is the total amplitude $\sqrt{|R_1|^2+|R_2|^2}$.}
\end{center}
\end{figure}

\section{Conclusions}\label{conclusions}

The main goal of this paper was to establish the inverse scattering formulae for the VNLS model on the half-line with integrable boundary conditions.
We had first to identify such integrable boundary conditions and this was done using the B\"acklund matrix approach. An important difference with 
the scalar case is the need to consider the time part of the B\"acklund transformation. Indeed, for the scalar case, this does not lead to new information 
and, in fact, it was not considered this way in the original papers \cite{BT1,BT2}. Only the space part was considered and then 
it was checked that the solution was compatible with time evolution. Here, we considered the time part directly and found that the solution to the space part 
has to be further 
reduced to two classes of possible integrable boundary conditions (which are then automatically compatible with the time evolution). On the technical side, 
this is chiefly 
due to the fact that in the scalar case, there is no difference between $RR^\dagger$ and $R^\dagger R$ while in the vector case, the former is a 
square matrix of size $n$ while the latter is a scalar. 

There seems to be a sharp contrast between vector and scalar cases in terms of which boundary conditions holding 
in the linear case go through to the interacting case. Indeed, in the scalar case, the Robin boundary condition $q_x(0,t)+dq(0,t)=0$, 
easily established for the free Schr\"odinger equation survives the switching on of the interaction in the sense that it is compatible 
with the integrability of the nonlinear Schr\"odinger equation. But here in the vector case, only subfamilies of the general Robin condition 
$R_{jx}(0,t)+d_jR_j(0,t)=0$, $j=1,\dots,n$ survive the requirement of integrability of the nonlinear equation when implemented in the 
form of the B\"acklund transformation method. 
Interestingly, this is not seen if one only looks at the 
$x$-part of the B\"acklund transformation method and the analysis of $t$-part also yields an important 
connection with the notion of linearizable boundary conditions in the Fokas method. 
In the Appendix, we examine in more details the linear and nonlinear systems.

Using the boundary conditions thus found, we implemented the vector version of the mirror image technique \cite{BH} to obtain 
the reconstruction formulae for the fields. Here again, a major difference with the scalar case is the appearance of coupled equations to determine 
the norming constants of the 
"virtual" solitons from the given scattering data characterizing the "real" solitons. We solved this problem explicitely in the case of one soliton.
Using these results, we discovered an interesting phenomenon: the boundary can be tuned to alter the polarization of a soliton. This was demonstrated 
with numerical simulations. 

The question of an efficient treatment of the general N-soliton problem in the presence of a boundary is an exciting open issue.

\section*{Acknowledgements}
We thank the referees for helpful comments and stimulating questions.

\section*{Appendix: Similarities and differences between linear and nonlinear Schr\"odinger equations with a boundary}

We take some time to discuss the classes of integrable boundary conditions that we have found using the B\"acklund transformation method. It is well known that 
for the free vector Schr\"odinger equation on the half-line (corresponding to switching off the nonlinearity by setting $\lambda=0$), the following  general
boundary conditions hold
\bea
\label{general_BC}
(U-I_n)R(0,t)+i(U+I_n)R_x(0,t)=0\,,
\eea
where $U$ is some $n\times n$ unitary matrix. This is the same as (\ref{general_Robin}) (and hence (\ref{general_Robin_diag}) in the boundary basis) 
where we allow formally some of the eigenvalues of $H$ to be infinite if the corresponding eigenvalues of $U+I_n$ are zero. But with the 
nonlinear term, we found that only subclasses of these general boundary conditions survive if we want to maintain integrability. Therefore, it appears 
that the nonlinearity imposes severe restrictions. One can then wonder if such restrictions are due to the use of the B\"acklund transformation method 
and if they could be circumvented by a generalization of it. 
There is an indication that the restrictions are genuine or at least consistent with existing methods. 
Indeed, in the quantum version of the system considered here,
the integrable boundary conditions are controlled by the reflection equation \cite{Skly_boundary}. The solutions for the corresponding reflection 
matrix have been classified in \cite{MRS2} and the same reduction of possible boundary conditions appears due to the nonlinearity. 
Algebraically, this means 
that the $R$ matrix of the Yangian of $gl(n)$ appears in the reflection equation instead of just the identity matrix in the free case.

To conclude this discussion, it is instructive to perform an analysis of the free case along the line of the Lax pair approach and compare with 
the nonlinear case. This sheds some more light on the differences between the two cases and on why integrability of the nonlinear case is likely to 
impose severe restrictions on the allowed boundary conditions.
In doing so, we provide an alternative and more direct proof of proposition \ref{prop_generating} in the mixed Neumann/Dirichlet case. In particular, 
this proof does not involve the B\"acklund matrix $L$.

Keeping the notations of section \ref{sect_Lax}, a Lax pair formulation for the free vector Schr\"odinger equation
\bea
iQ_t-Q_{xx}\Sigma_3=0\,,
\eea
reads
\bea
&&\mu_x + ik [ \Sigma _3 , \mu] =  Q \,,\\
&&\mu_t + 2 i k^2 [\Sigma_3, \mu]= 2k Q -  i Q_x \Sigma_3\,.
\eea
Splitting as usual in natural blocks, the linear analog of $\Gamma$ in proposition \ref{prop_generating} is $\mu_{21}\equiv \Gamma^0$ which satisfies the 
following linear version of the Ricatti equation (\ref{Ricatti})
\bea
\label{linear_Ricatti}
\Gamma^0_x=2ik\Gamma^0+R^\dagger\,,
\eea
which gives a series expansion $\displaystyle\Gamma^0=\sum_{n=1}^\infty\frac{\Gamma^0_n}{(2ik)^n}$ with $\Gamma^0_n=-\partial_x^{n-1} R^\dagger$.
We also have $(\Gamma^0 R)_t=i(\Gamma^0 R_x-\Gamma^0_xR)_x$ and hence, on the half-line, a generating function for the conserved charges is 
\bea
I^0(k)=\int_0^\infty\left(\Gamma^0(x,t,k)-\Gamma^0(x,t,-k)\right)R(x,t)\,dx\,,
\eea
with 
\bea
\label{linear_generating}
\partial_t I^0(k)&=&i\left(\Gamma^0_x(0,t,k)-\Gamma^0_x(0,t,-k)\right)R(0,t)-i
\left(\Gamma^0(0,t,k)-\Gamma^0(0,t,-k)\right)R_x(0,t)\,.\nonumber\\
\eea
Collecting everything, the charges are 
\bea
I^0_p=\int_0^\infty\partial_x^{2p}R^\dagger R\,dx~~,~~p\ge 0\,,
\eea
and they are conserved iff
\bea
\label{condition_p}
\left(\partial_x^{2p+1}R^\dagger R-\partial_x^{2p}R^\dagger \partial_x R\right)_{x=0}=0~~,~~p\ge 0\,.
\eea
Looking at $p=0$, the usual charge $I^0_0$ is conserved when the current is zero at $x=0$: $\left(R_x^\dagger R-R^\dagger R_x\right)_{x=0}=0$. 
This is equivalent to (\ref{general_BC}) and this is how these boundary conditions arise in the first place (note that they also ensure 
that the energy is real). It is rather easy to adapt the Fourier 
transform method to incorporate these boundary conditions. The beauty of the linear case is that then, all the higher charges are automatically
conserved (this is precisely the main difference with the nonlinear case).  Indeed, 
the general solution to the free Sch\"odinger equation with boundary conditions (\ref{general_BC}) can be written as
\bea
\label{free_field}
R(x,t)=\int_{\infty}^\infty\frac{dk}{2\pi}e^{ikx-ik^2t}A(k)\,,
\eea
where $A$ is a vector-valued function with the property $A(k)=B(-k)A(-k)$ where 
\bea
B(k)=-\left[(U-I_n)+k(U+I_n)\right]^{-1}\left[(U-I_n)-k(U+I_n)\right]\,.
\eea
But then, we get more than just (\ref{general_BC}). In fact, the reconstructed field (\ref{free_field}) satisfies the following mirror image symmetry
\bea
\label{mirror_property}
\left[(U-I_n)R+i(U+I_n)R_x\right](x,t)=-\left[(U-I_n)R+i(U+I_n)R_x\right](-x,t)\,,
\eea
of which (\ref{general_BC}) is a consequence. Hence, 
\begin{equation}
\left[(U-I_n)\partial_x^{2p}R+i(U+I_n)\partial_x^{2p+1}R\right](x,t)=-\left[(U-I_n)\partial_x^{2p}R+i(U+I_n)\partial_x^{2p+1}R\right](-x,t)\,,
\end{equation}
which immediately implies (\ref{condition_p}) as required.

Now for the nonlinear case, using the ingredients in the proof of proposition \ref{prop_generating}, we have 
\bea
\label{nonlinear_generating}
\partial_t\int_0^\infty(\Gamma(k)-\Gamma(-k))R\,dx&=&\left[i\left(\Gamma_x(k)-\Gamma_x(-k)\right)R-i
\left(\Gamma(k)-\Gamma(-k)\right)R_x \right.\nonumber\\
&&+\left.i(\Gamma(k)R\Gamma(k)R-\Gamma(-k)R\Gamma(-k)R)\right]_{x=0}\,,\nonumber\\
\eea
with 
\bea
\Gamma_x=2ik\Gamma+\lambda R^\dagger-\Gamma R\Gamma\,.
\eea
Comparing with (\ref{linear_generating}) and (\ref{linear_Ricatti}), the nonlinearity shows up at two levels: first there is an extra term in the 
right hand side of (\ref{nonlinear_generating}) and second, $\Gamma$ satisfies a nonlinear equation with the extra $\Gamma R\Gamma$ term compared to $\Gamma^0$.
Despite this fact, an explicit calculation shows that for the first charge, everything is the same as in the linear case. So one is led to impose 
(\ref{general_BC}). A crucial difference now is that one cannot use the Fourier transform but has to use the inverse scattering method instead. 
It is quite plain that this is likely to impose further restrictions on the possible integrable boundary conditions. The proof of proposition
\ref{prop_generating} shows that the B\"acklund transformation method selects such integrable boundary conditions. Although it is not a general proof, 
a careful inspection of the first few higher conserved charges (already for $n=2$) shows that there is little hope that more general boundary conditions 
than the ones we found ensure the conservation of all higher charges. In contrast, we now show that for the mixed Neumann/Dirichlet conditions, 
a nice argument 
similar to that of the linear case can be built to conclude that the left-hand side of (\ref{nonlinear_generating}) vanishes. This provides a
direct check of proposition \ref{prop_generating} and is based 
on the result of proposition \ref{mixed_symmetry} \ie $R(-x)=BR(x)$. The latter fact implies the important relation mirror image relation
\bea
\Gamma(x,t,k)=-\Gamma(-x,t,-k)B\,,
\eea
which is in fact the nonlinear analog of $R^\dagger(-x)=R^\dagger(x)B$ together will all the higher derivatives of this relation. As a consequence, 
$\Gamma(0,t,k)R(0,t)=-\Gamma(0,t,-k)R(0,t)$, $\Gamma(0,t,k)R_x(0,t)=\Gamma(0,t,-k)R_x(0,t)$ and $\Gamma_x(0,t,k)R(0,t)=\Gamma_x(0,t,-k)R(0,t)$. 
Inserting in the 
right-hand side of (\ref{nonlinear_generating}) yields zero as required.


\end{document}